\documentclass[a4paper,11pt]{article}
\usepackage{pos}
\usepackage{tikz}
\usepackage{CJKutf8}   

\title{Nucleon Axial Form Factor from Domain Wall on HISQ}

\author*[a,b]{Aaron S. Meyer}
\author[c,d]{Evan Berkowitz}
\author[e]{Chris Bouchard}
\author[f,a,b]{Chia Cheng Chang}
\author[g]{M.A. Clark}
\author[b]{Ben H\"orz}
\author[h,b]{Dean Howarth}
\author[i]{Christopher K\"orber}
\author[j]{Henry Monge-Camacho}
\author[k]{Amy Nicholson}
\author[l,m,f]{Enrico Rinaldi}
\author[h,b]{Pavlos Vranas}
\author[b,a,h]{Andr\'e Walker-Loud}

\affiliation[a]{Department of Physics, University of California,
  Berkeley, CA, USA}

\affiliation[b]{Nuclear Science Division, Lawrence Berkeley National Laboratory,
  Berkeley, CA, USA}

\affiliation[c]{Department of Physics, University of Maryland, College Park, MD 20742, USA}

\affiliation[d]{Institut f\"ur Kernphysik and Institute for Advanced Simulation, Forschungszentrum J\"ulich
54245 J\"ulich Germany}

\affiliation[e]{School of Physics and Astronomy, University of Glasgow, Glasgow G12 8QQ, UK}

\affiliation[f]{iTHEMS, RIKEN, 2-1 Hirosawa, Wako, Saitama 351-0198, Japan}

\affiliation[g]{NVIDIA Corporation, 2701 San Tomas Expressway, Santa Clara, CA 95050, USA}

\affiliation[h]{Physics Division, Lawrence Livermore National Laboratory, Livermore, CA 94550, USA}

\affiliation[i]{Institut f\"ur Theoretische Physik II, Ruhr-Universit\"at Bochum, D-44780 Bochum, Germany}

\affiliation[j]{Escuela de F\'isica , Universidad de Costa Rica, San Jos\'e, San Pedro, 11501, Costa Rica}

\affiliation[k]{Department of Physics and Astronomy, University of North Carolina, Chapel Hill, NC 27516-3255, USA}

\affiliation[l]{Physics Department, University of Michigan, Ann Arbor, MI 48109, USA}

\affiliation[m]{Theoretical Quantum Physics Laboratory, RIKEN, 2-1 Hirosawa, Wako, Saitama 351-0198, Japan}

\emailAdd{asmeyer.physics@gmail.com}

\abstract{
The Deep Underground Neutrino Experiment (DUNE) is an upcoming
 neutrino oscillation experiment that is poised to answer key questions
 about the nature of neutrinos.
Lattice QCD has the ability to make significant impact upon DUNE, beginning with computations of nucleon-neutrino interactions with \textit{weak} currents.
Nucleon amplitudes involving the axial form factor are part of the
 primary signal measurement process for DUNE,
 and precise calculations from LQCD can significantly reduce
 the uncertainty for inputs into Monte Carlo generators.
Recent calculations of the nucleon axial charge have demonstrated
 that sub-percent precision is possible on this vital quantity.
In these proceedings, we discuss preliminary results for the CalLat collaboration's
 calculation of the axial form factor of the nucleon.
These computations are performed with M\"obius domain wall valence quarks
 on HISQ sea quark ensembles generated by the MILC and CalLat collaborations.
The results use a variety of ensembles including several at
 physical pion mass.
}

\FullConference{%
 The 38th International Symposium on Lattice Field Theory, LATTICE2021
  26th-30th July, 2021
  Zoom/Gather@Massachusetts Institute of Technology
}

\begin{document}
\maketitle

\section{Introduction}

Within the next decade, several $O(\$1{\rm b})$ precision neutrino oscillation experiments
 will be coming online.
These experiments target as of yet unknown parameter values associated with
 neutrino flavor oscillation, such as the $CP$-violating phase
 and the neutrino mass hierarchy.
In addition, neutrino oscillation experiments can search for proton decay events
 and detect neutrinos from supernova sources.

The Deep Underground Neutrino Experiment (DUNE)~\cite{DUNE:2020mra}
 plans to measure neutrinos over the first oscillation maximum,
 which corresponds to an energy range of $1-10~{\rm GeV}$ for a baseline of 1300~km.
Many interaction topologies come into play in this energy range,
 with primary neutrino interaction classes including quasielastic scattering,
 resonance production, and deep inelastic scattering all in roughly equal proportions.
Quantification of the neutrino energy in events requires good control of cross sections
 for all of the relevant interaction topologies and sophisticated nuclear modeling
 to reconstruct the final statistical distributions and constrain oscillation parameters.

Of the interaction classes seen in neutrino oscillation experiments,
 quasielastic scattering is a primary signal measurement process due to its simplicity.
Quasielastic scattering dominates the total cross section at low neutrino energies,
 making it an important contribution for neutrino oscillation experiments such as
 HyperK, which will have neutrino energies strongly peaked in the quasielastic regime,
 and DUNE, which probes a large range of neutrino energies.
In this interaction class, a neutrino interacts with a
 freely propagating nucleon within a nucleus and becomes an outgoing charged lepton.
For these reasons, quasielastic scattering cross section amplitudes have
 stringent precision requirements and are a natural target for improving
 cross section systematics.

Since the neutrinos interact via a weak current,
 both vector and axial vector matrix elements are needed to compute the cross sections.
Unlike the vector matrix elements, the axial current contribution cannot be estimated
 from electron-proton scattering experiments.
Constraints on the axial matrix elements must come from either
 low statistics experimental measurements on elementary targets,
 model-dependent estimations from pion electroproduction,
 or large nuclear target neutrino scattering data with nuisance nuclear effects.

The uncertainty of the neutrino cross section amplitudes originating from
 nucleon form factors is large enough to be a possible cause of theory-experiment discrepancies.
Reanalysis of neutrino scattering data on elementary targets using the model-independent
 $z$ expansion reveals that the ubiquitous dipole model parameterization of the form
 factor underestimates the uncertainty by nearly an order of magnitude~\cite{Meyer:2016oeg}.
In the absence of a modern neutrino-deuterium scattering experiment,
 the most reasonable approach for quantifying and reducing uncertainties on the
 axial form factor is instead to compute nucleon matrix elements using lattice QCD and to feed
 the results into nuclear models.
Calculations are constructed to access the free nucleon form factor,
 completely circumventing the need for nuclear modeling.

\section{Simulation Details}

The utilized lattice action is the same as used by CalLat to compute $g_A$ with sub-percent precision~\cite{Chang:2018uxx,Berkowitz:2018gqe,Walker-Loud:2019cif}: a mixed lattice action with
M\"obius Domain Wall Fermions (MDWF) in the valence sector, solved
in a sea of Highly-Improved Staggered Quarks (HISQ)~\cite{Follana:2006rc}
and one-loop Symanzik improved gauge action
background field configurations.
The links used in the valence fermion Dirac operator were smeared with the gradient-flow scheme to a flow-time of $t_{gf}/a^2=1$~\cite{Berkowitz:2017opd}.  We are computing the axial form factor on approximately 30 ensembles, with seven pion masses ranging from $130\lesssim M_\pi \lesssim 400$~MeV, four lattice spacings of $0.06\lesssim a\lesssim0.15$~fm and multiple volumes~\cite{Miller:2020evg}.
In these proceedings, we present preliminary analysis on one ensemble, denoted a12m130, with $a\approx0.12$~fm and $M_\pi\approx130$~MeV, which was generated by the MILC collaboration~\cite{MILC:2012znn}.

Measurements are performed on 1000~configurations with 32 sources per configuration.
The two-point correlation functions are constructed from propagators
with point source quark fields smeared at both the source and sink
so that the correlation function is positive definite.
Additional quark propagators are sequentially solved with fixed source-sink time separations in the range $t/a\in\{3,...,12\}$, with sink-source spin polarizations up-up and down-down, for both forward time-propagating positive parity projectors and backwards time-propagating negative parity projectors.
The three-point correlation functions are constructed from the proper averaging of these four different correlation functions.
To reduce the cost of these sequential propagators, we use propagators from 8 sources to form a single coherent-sequential-sink~\cite{LHPC:2010jcs} as described in Ref.~\cite{He:2021yvm}.
The sink is projected to zero spatial momentum with the same quark smearing as at the source.
We then inject the current with all spatial momentum for all 16 quark bi-linear currents.
In this proceeding, we focus on the axial current aligned in the $z$ direction and the temporal vector current at 0 momentum (for normalization).

The axial form factor data are plotted as a ratio of a three-point over two-point
 correlation function.
The two-point functions are denoted $C^{\text{2pt}}(t,\mathbf{p})$
 with source-sink time $t$ and momentum $\mathbf{p}$, and
 three-point functions are denoted $C^{\text{3pt}}_{{\cal A}_z}(t,\tau,\mathbf{q})$
 with source-sink time $t$, current-insertion time $\tau$,
 insertion momentum transfer $\mathbf{q}$, and source momentum $-\mathbf{q}$.
Using these definitions, a correlator ratio that isolates the axial form factor is
\begin{align}\label{eq:ratio}
 {\cal R}_{{\cal A}_z}(t,\tau,\mathbf{q}) =&
 \frac{
   C^{\text{3pt}}_{{\cal A}_z}(t,\tau,\mathbf{q}) }{
   \sqrt{ C^{\text{2pt}}(t-\tau,\mathbf{0}) C^{\text{2pt}}(\tau,\mathbf{q}) }
 }
 \sqrt{
   \frac{ C^{\text{2pt}}(  \tau,\mathbf{0}) }{ C^{\text{2pt}}(t,\mathbf{0}) }
   \frac{ C^{\text{2pt}}(t-\tau,\mathbf{q}) }{ C^{\text{2pt}}(t,\mathbf{q}) }
}
\nonumber\\
\xrightarrow[t-\tau,\tau\to\infty]{}&
\frac{1}{\sqrt{2E_{\mathbf{q}} (E_{\mathbf{q}}+M)}}
\left[
 -\frac{q_z^2}{2M} \mathring{\tilde{g}}_P(Q^2)
 +(E_{\mathbf{q}}+M) \mathring{g}_A(Q^2)
\right]
\,,
\end{align}
where $\mathring{\tilde{g}}_{P}(Q^2)$ and $\mathring{{g}}_{A}(Q^2)$ are the respective unrenormalized induced pseudo scalar and axial form factors and $Q^2 = 2M^2(\sqrt{1+\frac{\mathbf{q}^2}{M^2}} -1)$.
In particular, for $q_z = 0$, the ground state of this ratio correlator is proportional to the unrenormalized axial form factor up to a computable kinematic factor.

In this analysis, a Bayesian framework is employed to fit the correlators.
The two-point and three-point correlation functions are fit with sums of exponentials with shared parameters,
\begin{align}
&C^{\text{2pt}}(t,\mathbf{p}) = \sum^N_n |z^{\mathbf{p}}_n|^2 e^{-E^{\mathbf{p}}_n t}\, ,&
&C^{\text{3pt}}_{{\cal A}_z}(t,\tau,\mathbf{q}) =
 \sum^N_{m,n} z^{\mathbf{0}}_n z^{\mathbf{q}}_m A^{\mathbf{q}}_{nm}
 e^{-E^{\mathbf{0}}_n (t-\tau)} e^{-E^{\mathbf{q}}_m \tau}\, ,&
\end{align}
with momentum $\mathbf{p}$ and current insertion momentum $\mathbf{q}$.

A tower of $N$ exponential contributions is included for each state,
 with the choice of $N=3$ for all correlators.
The simultaneous fit includes the 0-momentum temporal vector current
 in addition to the axial current in order to get the vector charge,
 which provides an extra constraint on the spectrum of states.
With our lattice action, $Z_A/Z_V -1 \leq 5\times10^{-5}$ for all ensembles~\cite{Chang:2018uxx} and so
the ratio $\mathring{g}_A(Q^2)/\mathring{g}_V$ is normalized absolutely.

The posterior values for the ground state-to-ground state transitions
 at various momentum transfers obtained from the correlation function fit
 are the desired nucleon axial form factor data.
The form factor parameterization of choice is the $z$ expansion~\cite{Meyer:2016oeg}
\begin{align}
 g_A(Q^2) = \sum_{k=0} a_k \big[ z(Q^2) \big]^k.
 \label{eq:gazexp}
\end{align}
with a conformal mapping of the form,
\begin{align}
 z(Q^2) = \frac{\sqrt{t_c +Q^2} -\sqrt{t_c -t_0}}{\sqrt{t_c +Q^2} +\sqrt{t_c -t_0}}.
 \label{eq:zdef}
\end{align}
The parameter $t_c$ is a kinematic cutoff ($t_c=9M_\pi^2$ for the axial current)
 and $t_0$ is a free parameter that may be set to improve convergence.

The form in Eq.~(\ref{eq:zdef}) ensures that $|z|<1$ for quasielastic scattering
 so that the expansion parameter $z$ that appears in Eq.~(\ref{eq:gazexp})
 is guaranteed to be small.
The sum in Eq.~(\ref{eq:gazexp}) is in practice truncated at finite order $k_{\rm max}$,
 so the large-$Q^2$ behavior is controlled by including extra parameters
 for $k=k_{\rm max}+n+1$ and enforcing sum rules of the form
\begin{align}
 \biggr( \frac{\partial}{\partial z} \biggr)^{n}
 \; \sum_{k=0}^{k_{\rm max}+4} a_k z^k \Big|_{z=1} = 0
 \label{eq:sumrules}
\end{align}
 with $n\in\{0,1,2,3\}$.
The coefficients $a_k$ in the form factor fit are given priors
\begin{align}
 {\rm prior}\biggr[ \frac{a_k}{|a_0|} \biggr] = 0 \pm {\rm min} \biggr[ 5, \frac{25}{k} \biggr]
\end{align}
 as done in Ref.~\cite{Meyer:2016oeg}.

\section{Results}

The ranges of the fit time have been chosen such that the minimum time,
 $t_{\rm min}$, is consistent across both the two-point and three-point functions:
 the two-point correlators are fit to the time range $t \in [t_{\rm min}, t_{\rm max,2}]$
 and the three-point correlators with source-sink separation $t$ to the range
 $\tau \in [t_{\rm min}, t-t_{\rm min}]$.
This choice ensures that the minimum time separation between any two operator insertions
 is at least $t_{\rm min}/a$ timeslices apart.
All available three-point data that satisfy these temporal restrictions are included in the fit.
All discrete 3-momenta that satisfy
 $|q_{x,y}| \leq 4\sqrt{2}\cdot (2\pi/L)$ and $q_z=0$
 are simultaneously fit to extract posteriors,
 which corresponds to four-momentum transfers up to about $1.06~{\rm GeV}$~\cite{Miller:2020evg}.

\newcommand{\pltwid}{0.3\textwidth}
\begin{figure}[tbh!]
\begin{tikzpicture}
\node at (0,0) {
\setlength\tabcolsep{-1em}
\begin{tabular}{cccc}
\includegraphics[width=\pltwid]{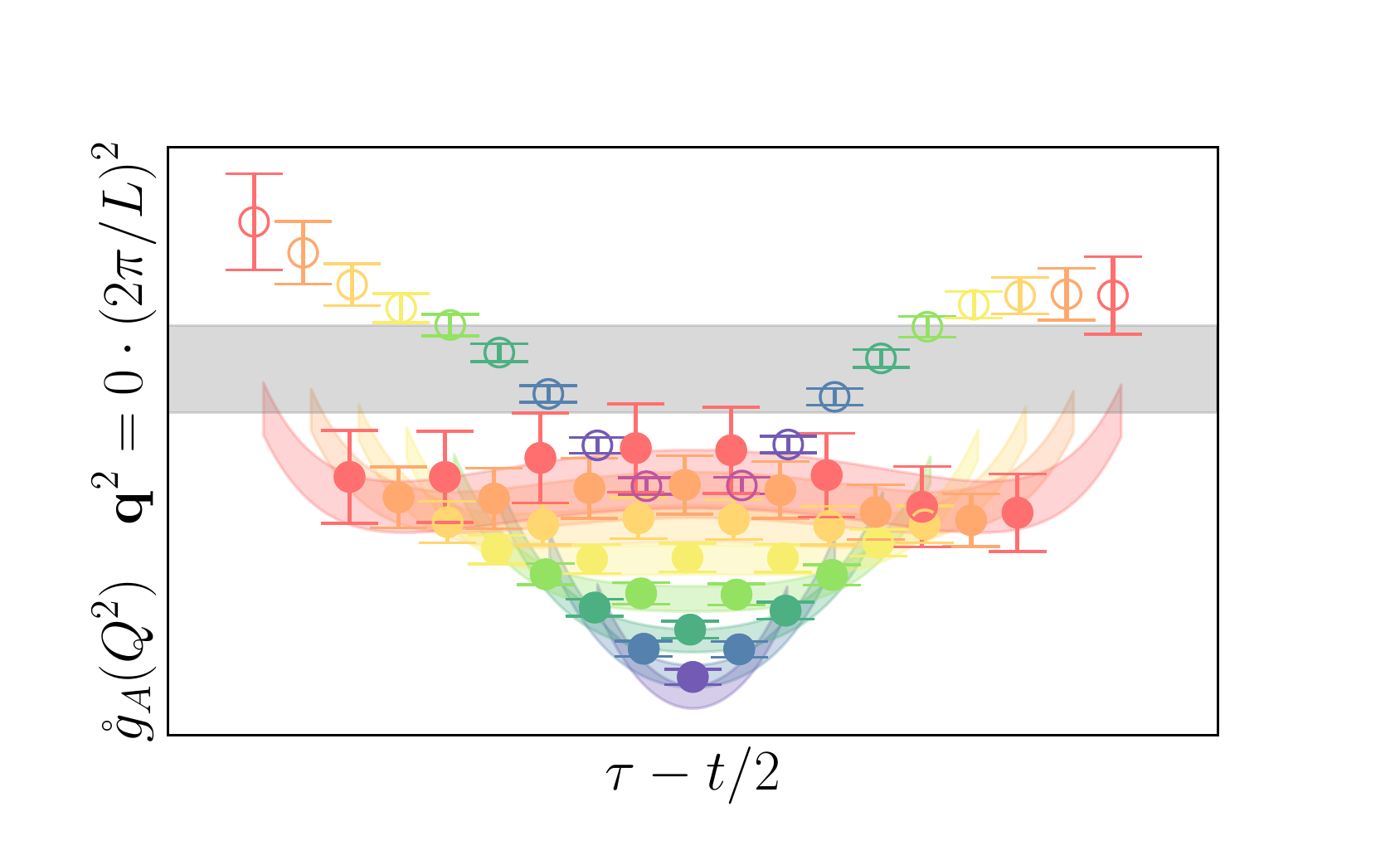} &
\includegraphics[width=\pltwid]{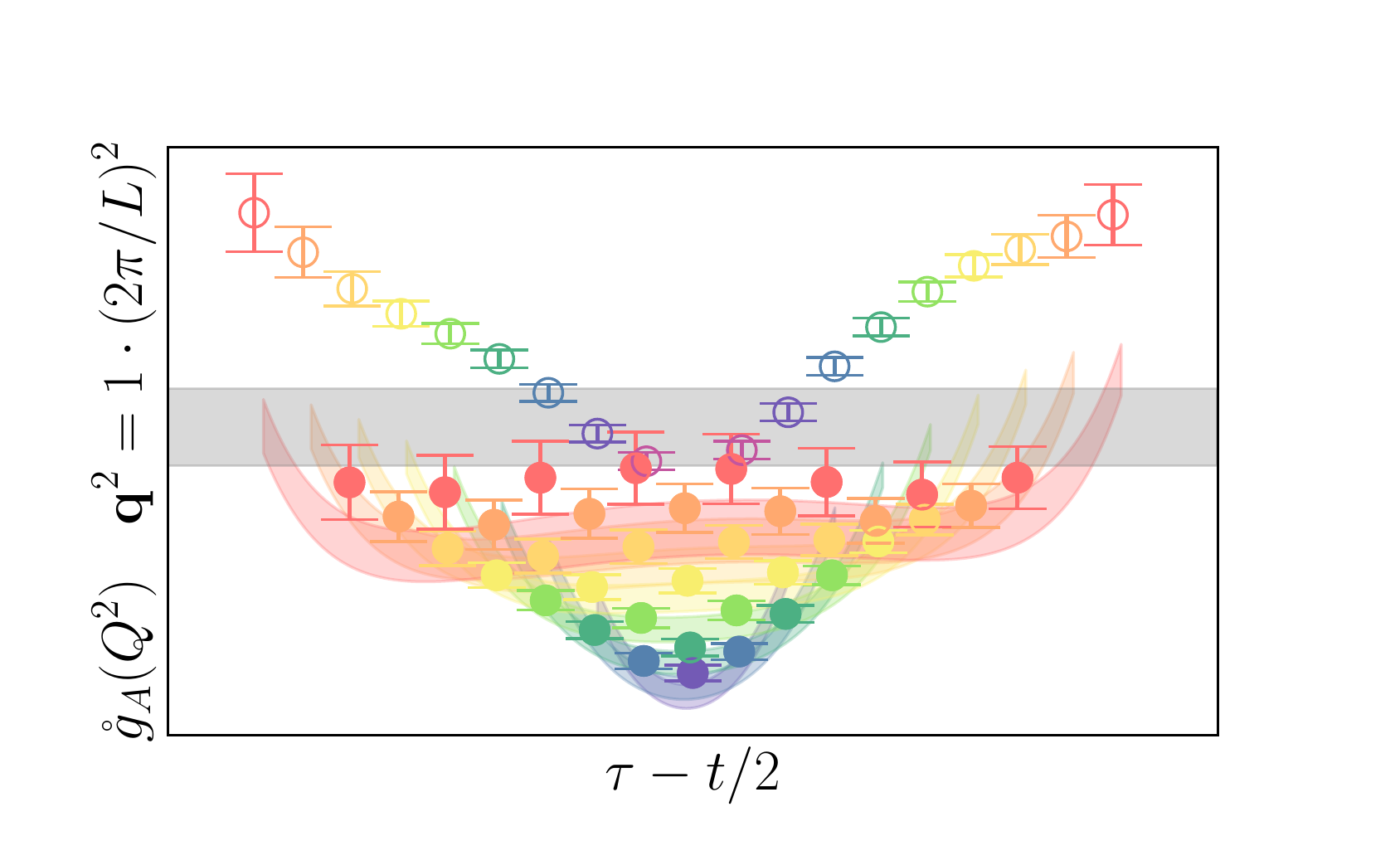} &
\includegraphics[width=\pltwid]{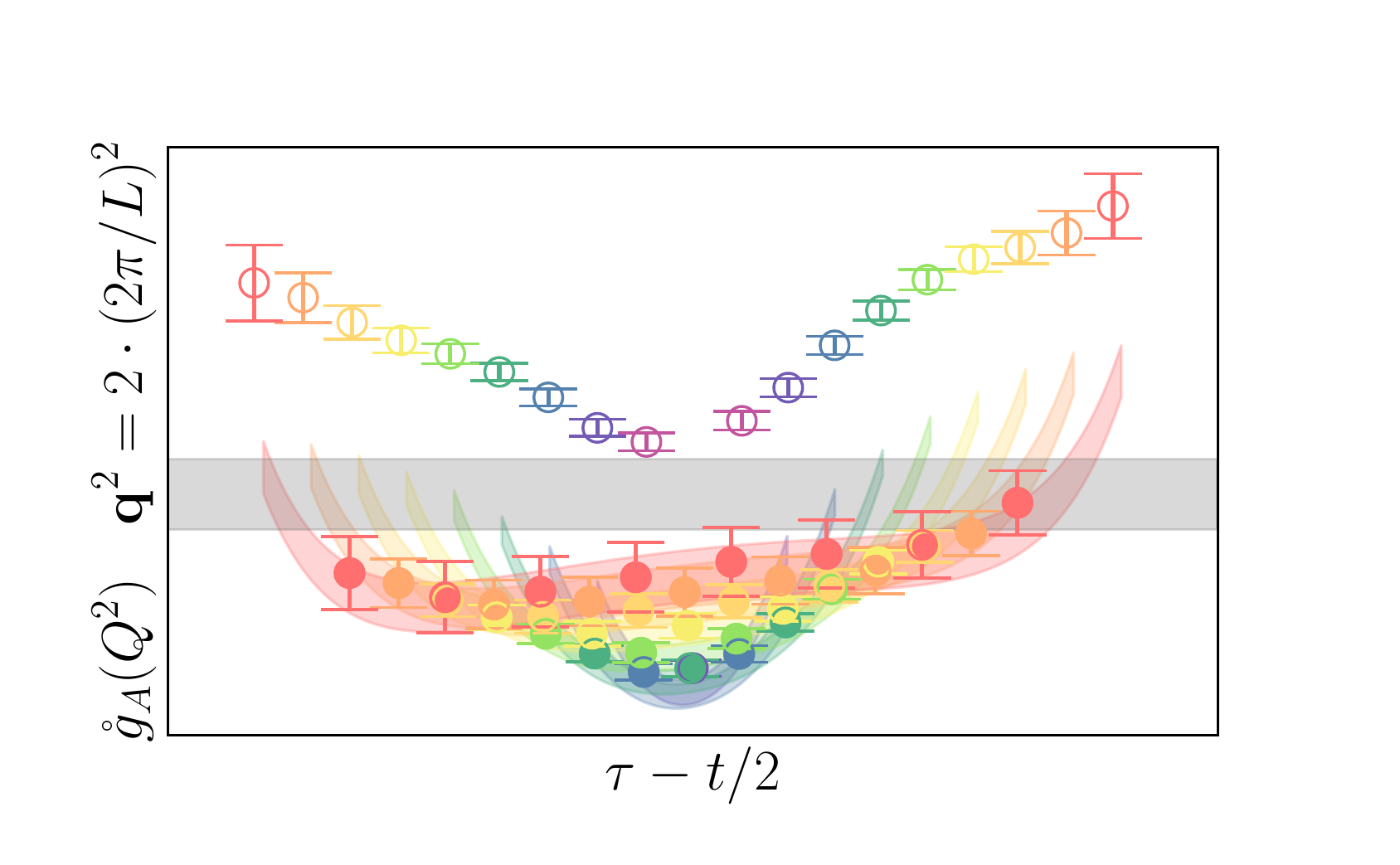} &
\includegraphics[width=\pltwid]{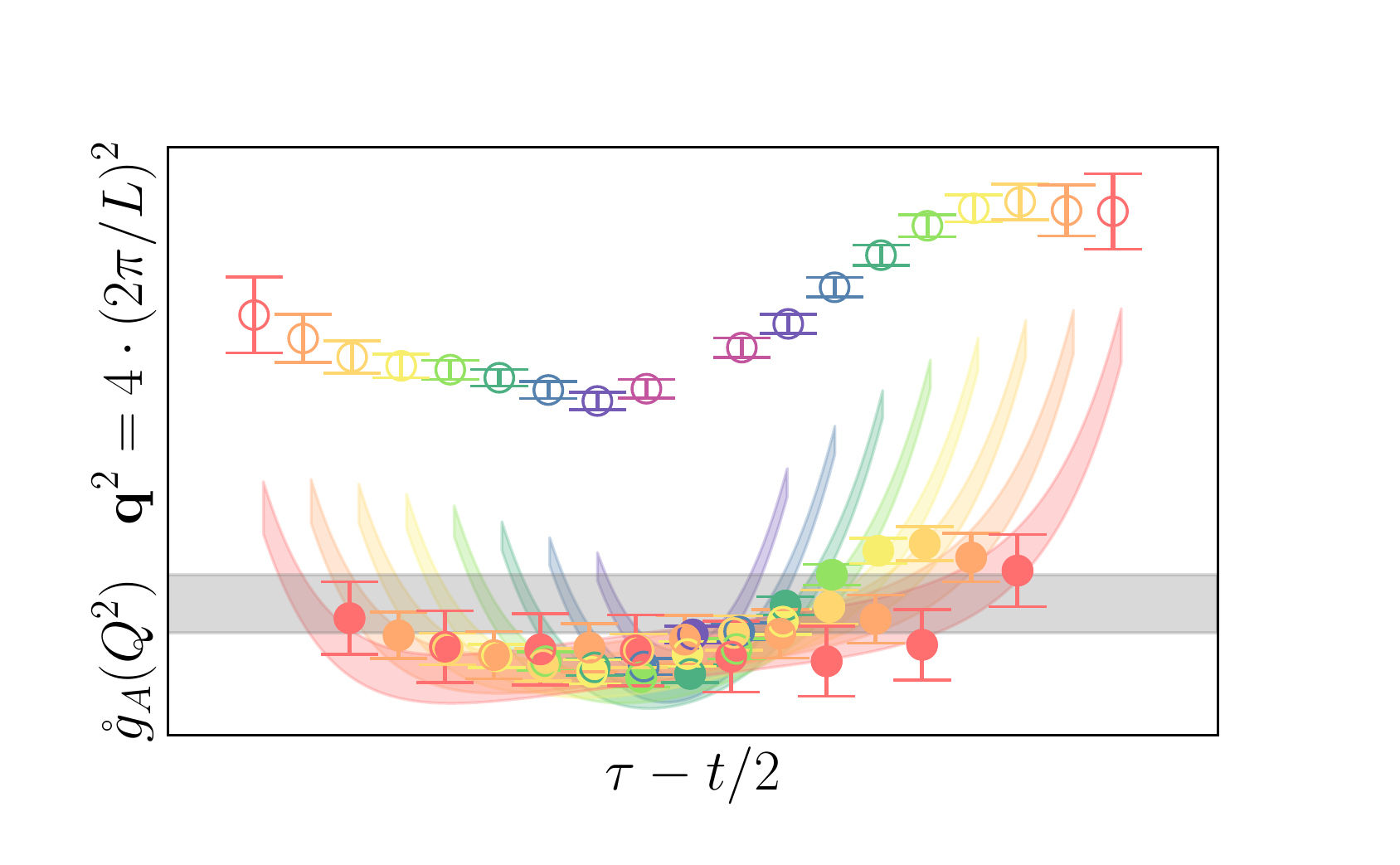} \\[-2em]
\includegraphics[width=\pltwid]{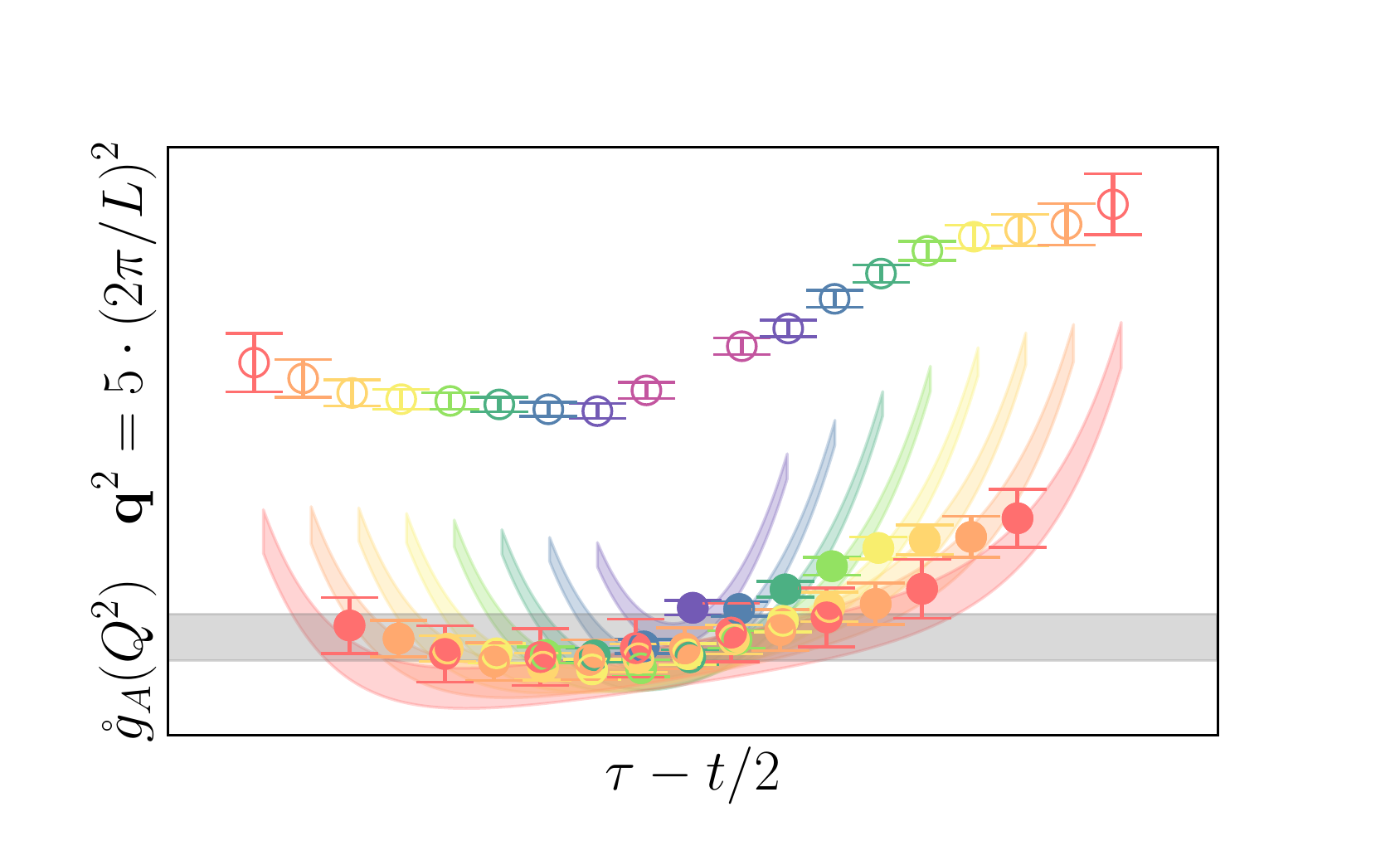} &
\includegraphics[width=\pltwid]{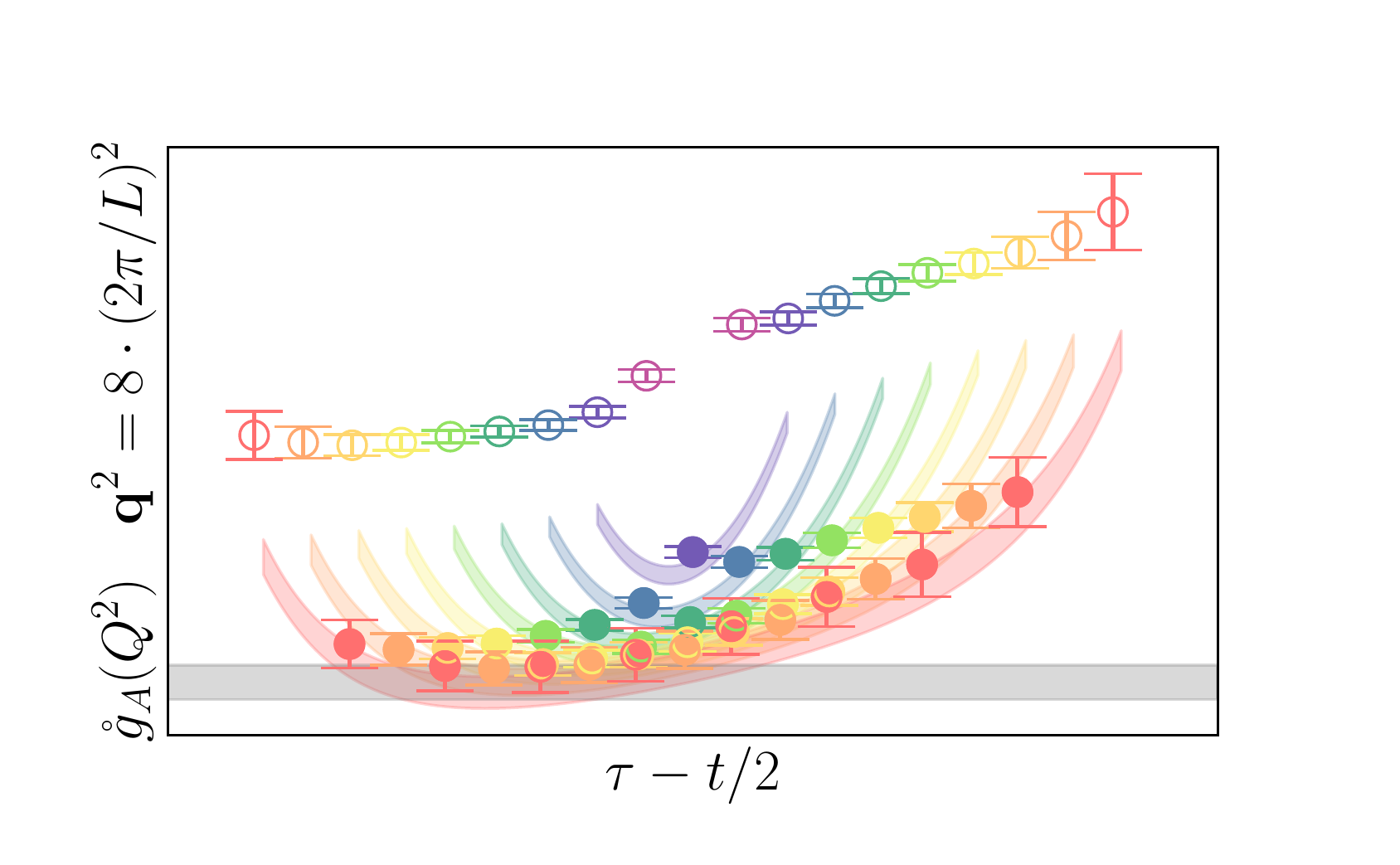} &
\includegraphics[width=\pltwid]{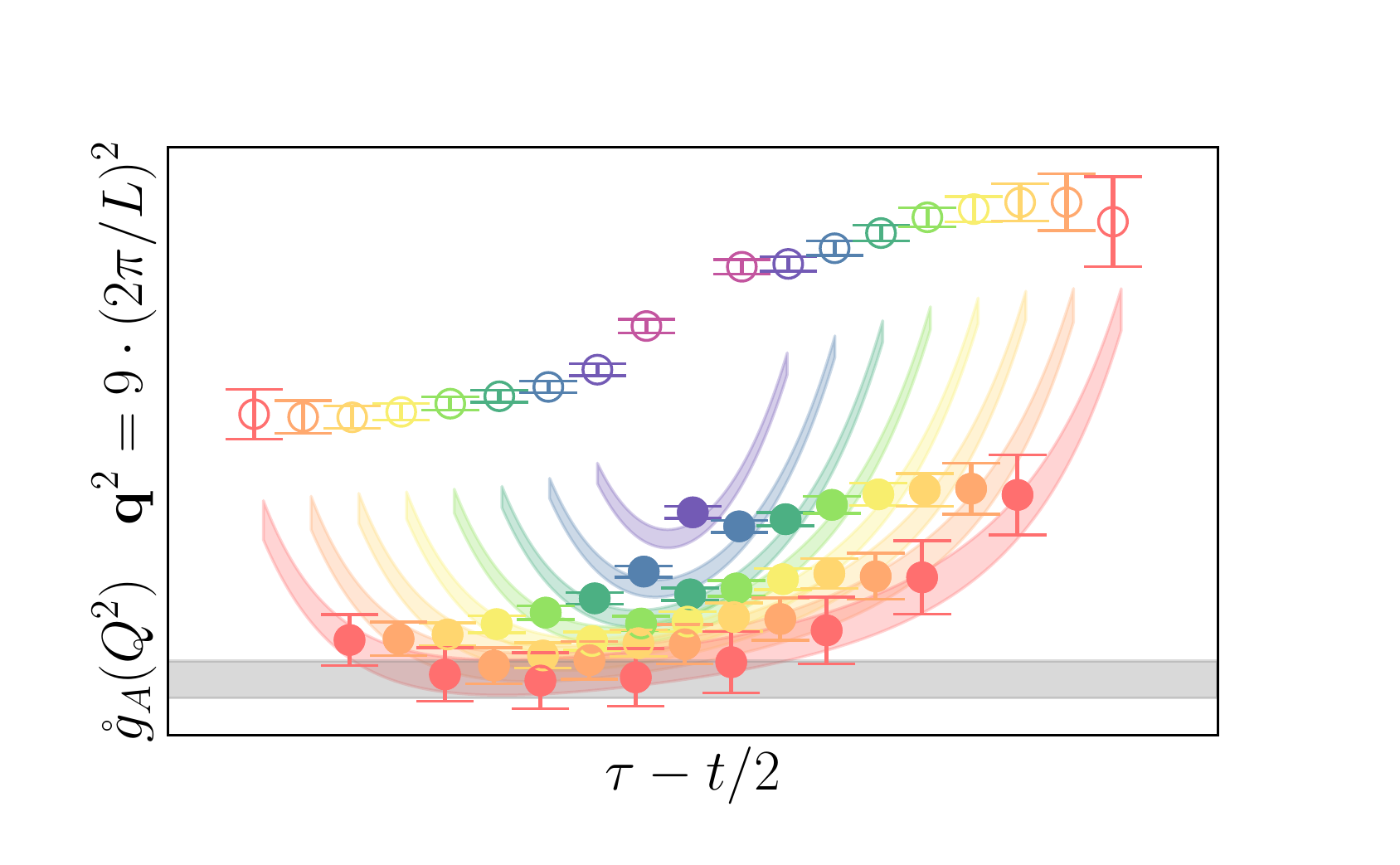} &
\includegraphics[width=\pltwid]{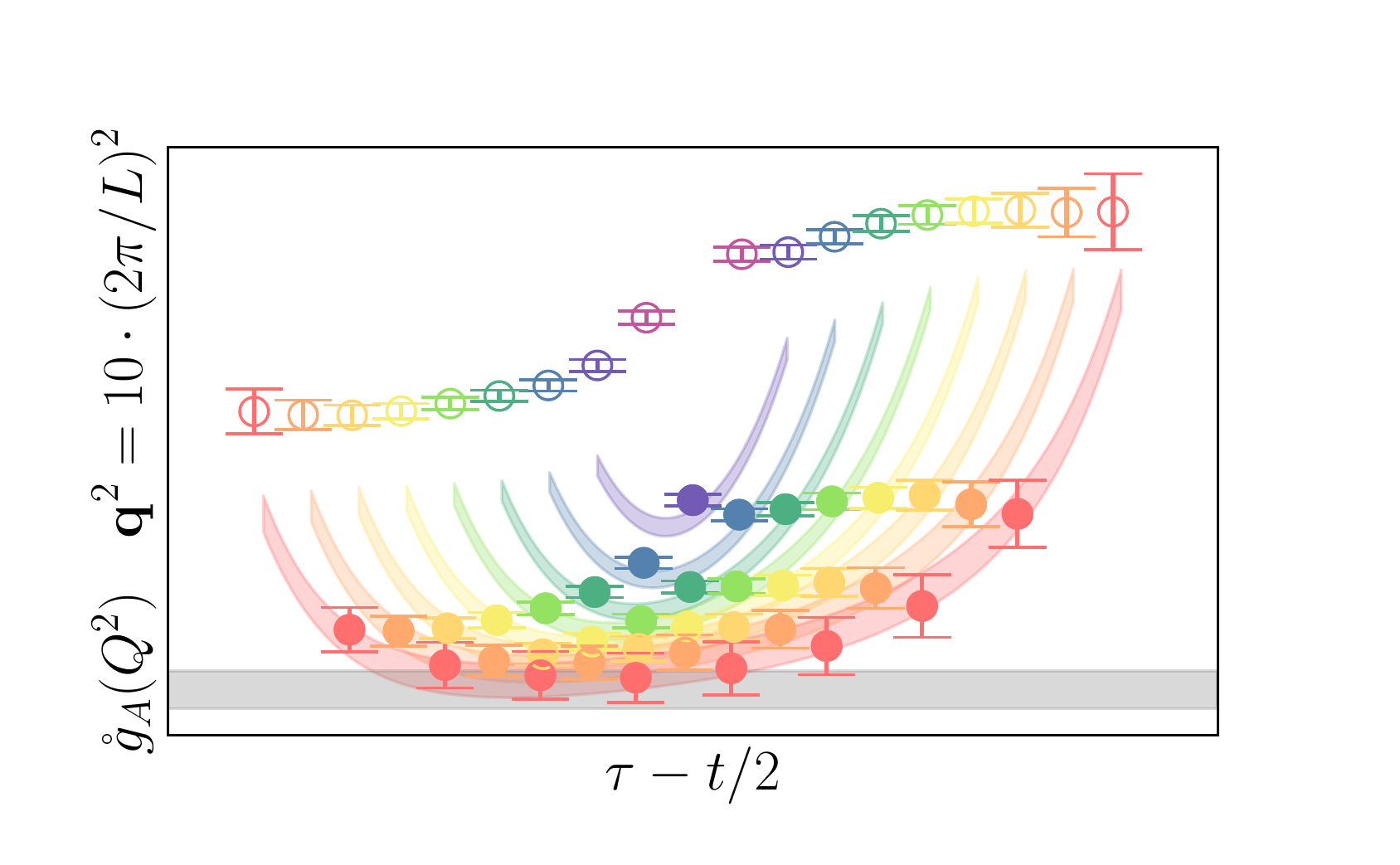} \\[-2em]
&
\includegraphics[width=\pltwid]{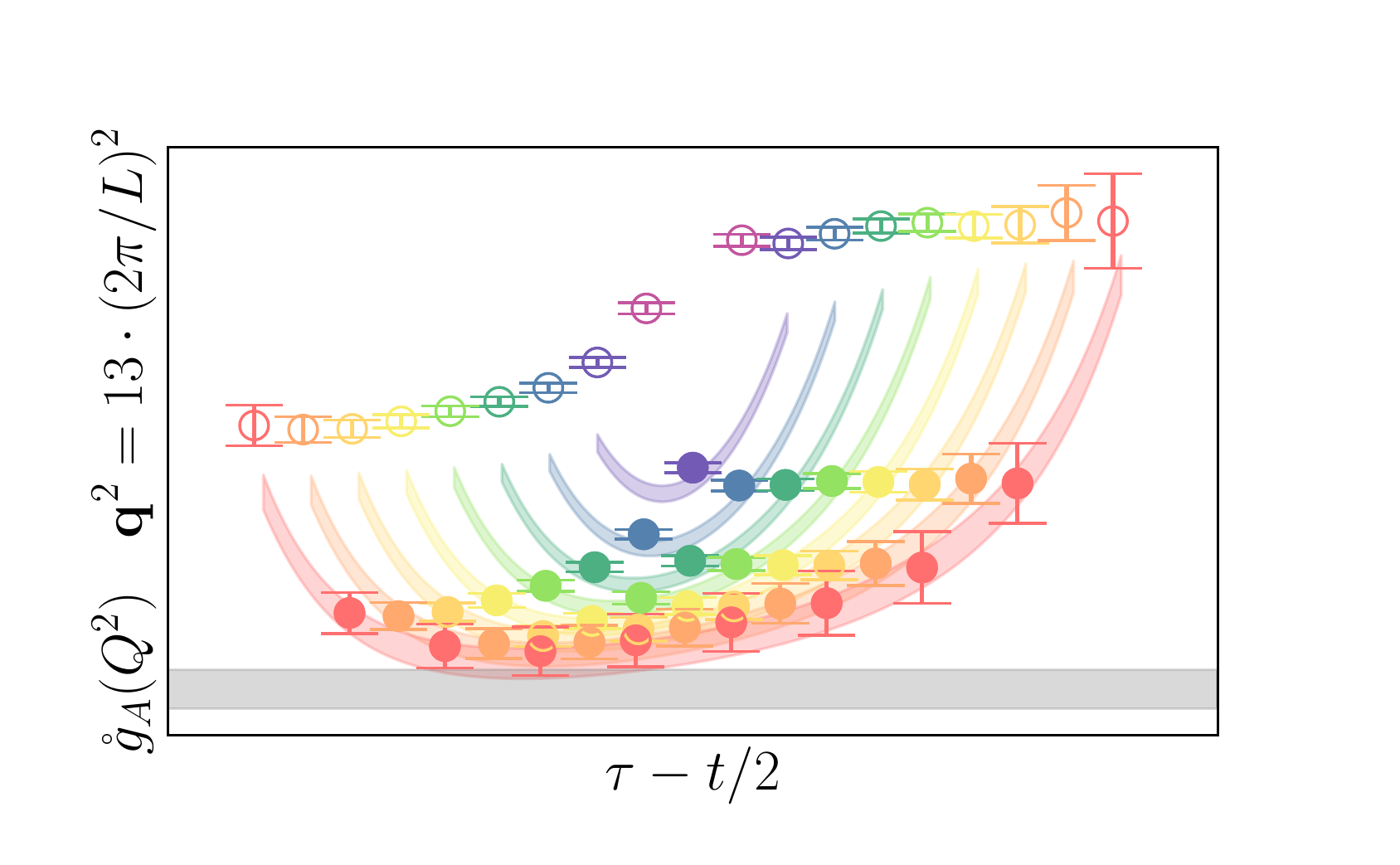} &
\includegraphics[width=\pltwid]{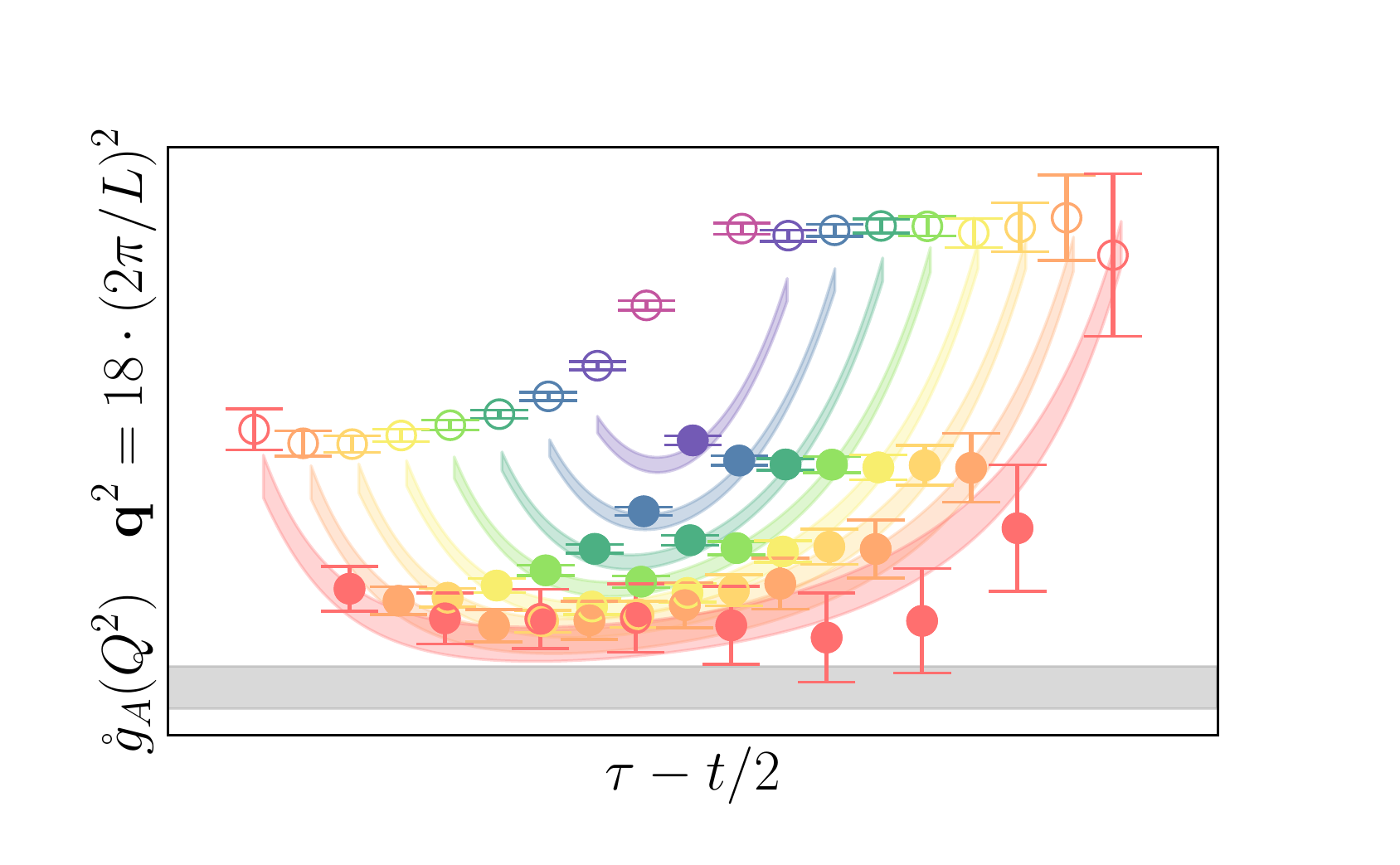} \\
\end{tabular}
 };
\node[opacity=0.1,gray] at (0,0) {\scalebox{5.4}[20]{PRELIMINARY}};
\end{tikzpicture}
\caption{
 Plot of the ratio in Eq.~(\ref{eq:ratio}) formed from the raw correlator data.
 The horizontal axis is the source-insertion separation time minus half
  the source-sink separation time to center the data.
 The gray band is the posterior value for $\mathring{g}_A(Q^2)$
  after a 3-exponential correlator fit, including all $m\rightarrow n$ transitions with $0\leq m,n<3$.
 The data used in the fit are shown as filled-in circles and data outside the fit range
  are shown as unfilled circles.
 Different source-sink separations are plotted as different colors,
  ranging from the shortest source-sink separation $t/a=3$ (purple)
  to the longest source-sink separation with $t/a=12$ (red).
 The posterior curves obtained from an exponential fit is plotted as the colored band,
  with the color corresponding to the same source-sink separation as the data.
 The different panels correspond to different 3-momentum transfers squared,
  in units of the minimum lattice momentum $2\pi/L$ squared, for the lowest 10 momenta used in the analysis.
 Since the data are preliminary, the units on the vertical axis are omitted.
 \label{fig:pdm}
}
\end{figure}

The data and posterior fits are combined with the ratio written in Eq.~(\ref{eq:ratio}),
 which is plotted in Fig.~\ref{fig:pdm} for a sub-set of the correlators.
This figure includes the lowest 10 momenta with $q_z=0$,
 and all of the available three-point data are plotted with their appropriate ratios.
The fits to the correlator data show broad agreement across all ratio values plotted,
 down to time separations as small as $t,\tau,(t-\tau)=2a$.

There are several qualities that suggest large excited-state contaminations at low-momentum transfers.
There is a strong curvature with $\tau/a$ at low momentum transfer,
 even for the largest source-sink time separation (red), with a curvature that changes sign.
The smallest source-sink time separations are relatively smaller than the largest
 source-sink time separations at low momentum, but the opposite is true at large momenta.
The gray band corresponding to the posterior matrix element connecting ground states
 of ingoing and outgoing momenta is as much as $2\sigma$ from the ratio central value.
These qualities are suggestive of large excited state contaminations
 in the low-momentum transfer data, particularly the axial charge.

As the momentum is increased, the agreement between the ground state posterior gray band
 and the ratio value of the largest source-sink separation come into agreement.
The ratio data for the largest source-sink separation times fall on top of each other,
 suggesting that the most harmful excited state contaminations have decayed away
 and only the ground state contribution remains.
These two observations give confidence that the excited states are reasonably
 controlled, and the posteriors give the axial matrix elements of interest.

The fit to the axial matrix elements is shown in Fig.~\ref{fig:gaq2}.
The green (lighter top) curve is a 5-parameter fit to the scatter points
 with a $z$ expansion parameterization including 4 sum rule constraints,
 which gives a good description of the data.
The gray (darker lower) curve is a 2-parameter dipole fit.
The precision of the data is particularly good,
 with roughly constant uncertainties for
 all of considered momentum transfers.

\renewcommand{\pltwid}{0.75\textwidth}
\begin{figure}[tbh!]
\centering
\begin{tikzpicture}
\node at (0,0) {
 \includegraphics[width=\pltwid]{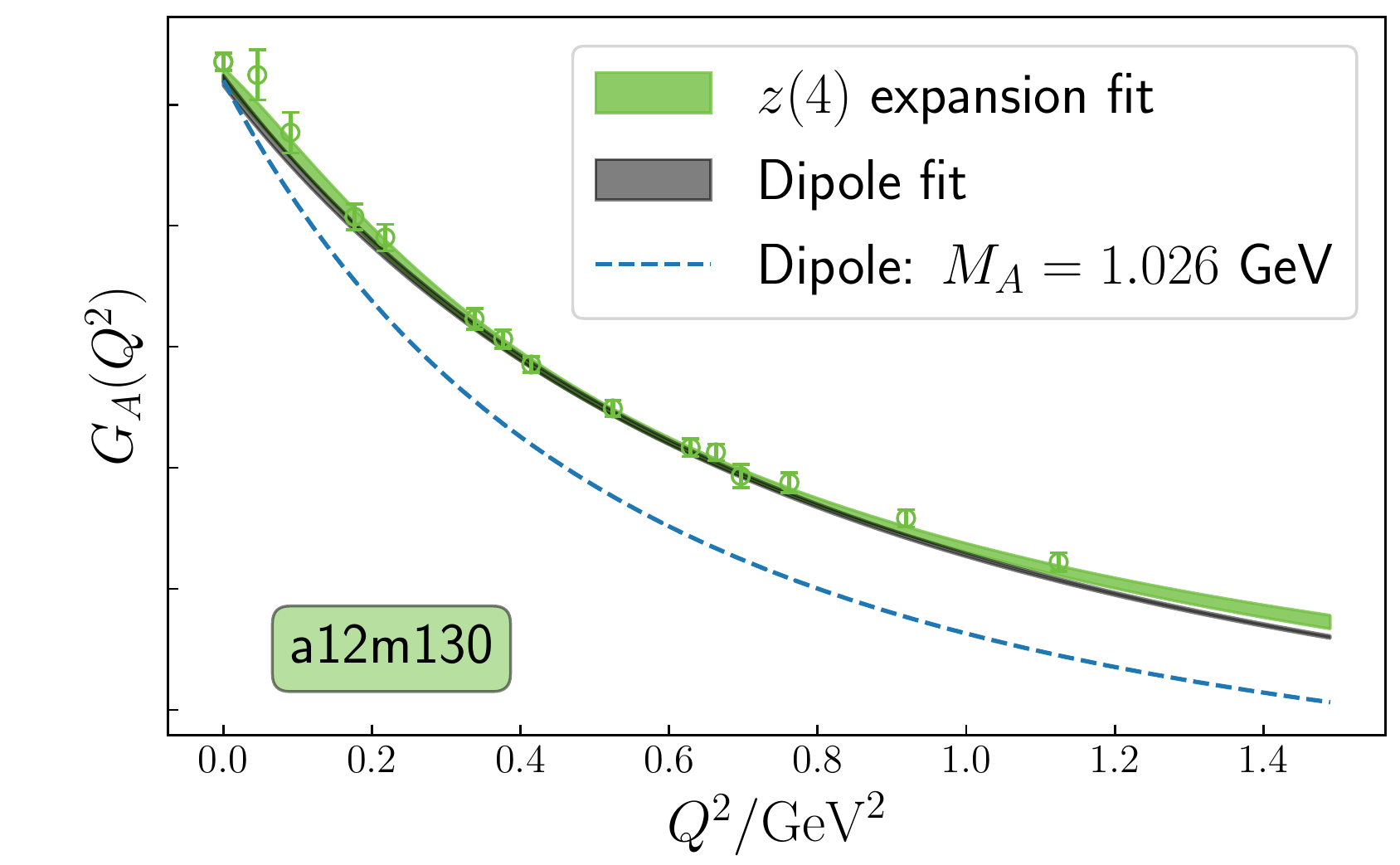}
 };
\node[red] at (-2.0,-0.8) {\scalebox{1.0}{PRELIMINARY}};
\end{tikzpicture}
\caption{
 Plot of the renormalized axial form factor as a function of $Q^2$.
 The scatter points are the posteriors from the form factor obtained after fitting
  to the correlation functions with their uncertainties.
 The upper (green) band is the $z$ expansion fit to the scatter points.
 The lower (gray) band is the dipole fit.
 The dashed line is a dipole parameterization with $M_A=1.026$~GeV~\cite{Bernard:2001rs}.
 Since the data are preliminary, the vertical axis scale is omitted
  and the lower axis bound is set to a nonzero value.
 \label{fig:gaq2}
}
\end{figure}

The form factor fits in this analysis achieve sub-percent precision on the form factor.
In particular, the axial charge is constrained with a relative precision of $0.6\%$, slightly better than  CalLat's axial charge analysis~\cite{Chang:2018uxx,Berkowitz:2018gqe,Walker-Loud:2019cif}.
The form factor has a similar absolute precision out to larger momentum transfers.
Taking the axial radius squared as a definitive metric for the axial form factor,
 with the definition
\begin{align}
 r_A^2 = -\frac{6}{g_A(0)}
 \frac{d g_A(Q^2)}{d Q^2} \Big|_{Q^2=0},
\end{align}
 a relative precision of $\delta r_A^2 / r_A^2 \approx 0.13$ is achieved,
 more than a factor of 3 more precise than the radius obtained from
 neutrino scattering on deuterium~\cite{Meyer:2016oeg}.
In addition, the axial radius obtained is smaller than observed in experiments.
This points to a slower falloff with $Q^2$ than expected,
 a trend that is consistent with other LQCD extractions.
This finding indicates that more weight should be given to larger
 momentum transfer neutrino interactions,
 which could change the relative frequency of quasielastic scattering
 relative to other interaction topologies.

\section{Outlook}

Lattice QCD has the potential to make an impact on the neutrino oscillation
 program by providing significantly more precise nucleon form factors
 for weak interaction amplitudes.
Existing analyses of form factor data have the potential to reach percent-level
 constraints on the form factors with a full error budget,
 an order of magnitude more precise than the axial form factor constraints
 obtained from neutrino scattering on elementary targets.
With this level of precision,
 and barring existing tensions in the vector form factor parameterizations~\cite{Borah:2020gte},
 nucleon-level form factor uncertainties can be made subdominant compared
 to nuclear modeling uncertainties,
 enabling robust inputs with realistic uncertainties for nuclear model calculations.

Excited states in the axial matrix elements have always been a difficulty
 for nucleon matrix elements involving an axial current.
Incomplete characterization of excited states are believed to have historically
 led to extractions of the axial charge that are low compared to experiments.
A significant contamination comes from transitions of the ground state nucleon
 to excited states, possibly including multiparticle states involving
 a nucleon and a pion.
Inclusion of many source-sink time separations down to short times
 has been shown to be beneficial for control of excited states,
 leading to more precise estimates of ground state
 matrix element information~\cite{He:2021yvm}.
This analysis has included at least 8 source-sink separation times over 10 momenta
 and has demonstrated reasonable control over a 3-state fit for all momentum combinations.

The axial form factor data on this single ensemble look very promising,
 with an expected percent-level precision.
Scaling of uncertainties with the momentum transfer is better than expected
 when considering 4-momentum transfers squared up to $0.7~{\rm GeV}^2$,
 which use only a small fraction of the available data.
Data for 3-momentum-squared up to 5 times larger than those considered are available,
 which would permit constraints on the form factor of $Q^2$ up to a few ${\rm GeV}^2$.
These data further feed back to better constraints on the ground state spectrum
 and matrix elements, providing strong constraints on the axial charge
 rest-frame spectrum.

\section{Acknowledgements}

This work was supported in part by
 the NVIDIA Corporation (MAC),
 the Alexander von Humboldt Foundation through a Feodor Lynen Research Fellowship (CK),
 the RIKEN Special Postdoctoral Researcher Program (ER),
 the Nuclear Physics Double Beta Decay Topical Collaboration (HMC, AN, AWL),
 the U.S. Department of Energy, Office of Science, Office of Nuclear Physics under Award
  Numbers DE-AC02-05CH11231 (CCC, CK, BH, AWL),
  DEAC52-07NA27344 (DH, PV),
  DE-FG02-93ER-40762 (EB),
  DE-SC00046548 (ASM);
  the DOE Early Career Award Program (AWL),
  and the U.K. Science and Technology Facilities Council grants ST/S005781/1 and ST/T000945/1 (CB).

Computing time for this work was provided through the
 Innovative and Novel Computational Impact on Theory and Experiment (INCITE) program
 and the LLNL Multi programmatic and Institutional Computing program
 for Grand Challenge allocations on the LLNL supercomputers.
This research utilized the NVIDIA GPU accelerated Summit supercomputer
 at Oak Ridge Leadership Computing Facility at the Oak Ridge National Laboratory,
 which is supported by the Office of Science of the
 U.S. Department of Energy under Contract No. DEAC05-00OR22725
 as well as the Lassen supercomputer at Lawrence Livermore National Laboratory.

 The computations were performed with \texttt{LaLiBe}~\cite{lalibe}, linked against \texttt{Chroma}~\cite{Edwards:2004sx} with \texttt{QUDA} solvers~\cite{Clark:2009wm,Babich:2011np} and HDF5~\cite{hdf5} for I/O~\cite{Kurth:2015mqa}.
 They were efficiently managed with \texttt{METAQ}~\cite{Berkowitz:2017vcp,Berkowitz:2017xna} and EspressoDB~\cite{Chang:2019khk}.
 The numerical analysis utilized \texttt{gvar}~\cite{gvar} and \texttt{lsqfit}~\cite{lsqfit}.

\bibliographystyle{JHEP}
\bibliography{main}

\end{document}